\documentclass[english]{article}
\usepackage[T1]{fontenc}
\usepackage[latin9]{inputenc}
\usepackage{geometry}
\usepackage{color}
\usepackage{array}
\usepackage{amsmath}
\usepackage{amssymb}
\usepackage{graphicx}
\usepackage{epstopdf}
\usepackage{epsfig}
\usepackage{esint}

\makeatletter

\providecommand{\tabularnewline}{\\}

\makeatother

\usepackage{babel}

\begin{document}

\title{A comparative study of protocols for secure quantum communication
under noisy environment: single-qubit-based protocols versus entangled-state-based protocols}

\author{Vishal Sharma, Kishore Thapliyal, Anirban Pathak\thanks{anirban.pathak@gmail.com},
Subhashish Banerjee }
\maketitle
\begin{abstract}
The effect of noise on various protocols
of secure quantum communication has been studied. Specifically, we have investigated the effect
of amplitude damping, phase damping, squeezed generalized amplitude
damping, Pauli type as well as various collective noise models
on the protocols of quantum key distribution, quantum key agreement,
quantum secure direct quantum communication and quantum dialogue.
From each type of protocol of secure quantum communication, we have
chosen two protocols for our comparative study; one based
on single qubit states and the other one on entangled
states. The comparative study reported here has revealed that single-qubit-based
schemes are generally found to perform better in the presence of amplitude
damping, phase damping, squeezed generalized amplitude damping noises,
while entanglement-based protocols turn out to be preferable in the
presence of collective noises. It is also observed that the effect
of noise entirely depends upon the number of rounds of quantum communication
involved in a scheme of quantum communication. Further, it is observed that squeezing, a completely quantum mechanical resource present in the
squeezed generalized amplitude channel, can be used in a beneficial way as it may yield  higher fidelity compared to the corresponding zero squeezing case.
\end{abstract}

\section{Introduction}

In 1984, Bennett and Brassard proposed the first protocol for quantum
key distribution (QKD) which is now known as BB84 protocol \cite{bb84}.
This pioneering work drew a considerable amount of attention from the scientific community,
as it was shown to be able to provide unconditional security, a desired
feature for all key distribution schemes, but unachievable in the
domain of classical cryptography. The fact that unconditional security
can be provided if we use quantum resources for key distribution led
to extensive studies on the protocols of secure quantum communication
(see Ref. \cite{my-book} for further details). Initial studies were limited
to QKD \cite{bb84,b92,ekert,BBM,vaidman-goldenberg}. These
initial studies on QKD  brought out a number of facts which were further
established later. Here, in the context of the present
work, we wish to specially stress on a specific aspect. In 1991,
Ekert proposed a protocol for QKD using entangled state which can
be reduced to BB84 protocol (which uses single photon (qubit) states)
under certain conditions \cite{ekert}. Later, Bennett introduced
a single-photon-based scheme for QKD which requires only 2 states, 
now known as the B92 protocol \cite{b92}. Soon BBM protocol
\cite{BBM} was introduced, and it was found that BBM protocol may
be viewed as an entangled-state-based analogue of the single-photon-based
B92 protocol. Thus, these studies indicated that the security achieved
by a single-photon-based scheme can also be achieved by a corresponding
entangled-state-based scheme. As we have already mentioned, initial studies
on quantum cryptography were limited to QKD \cite{bb84,b92,ekert,BBM,vaidman-goldenberg}.
Later on, several other aspects of secure quantum communication were
investigated. For example, protocols were proposed
for quantum secret sharing \cite{Hillery}, quantum key agreement
(QKA) \cite{QKA,Qka-CS,QKA-singlepar}, quantum dialogue (QD) \cite{ba-an,baan_new,qd,QD-singlepar,QD-hwang}, quantum secure direct communication (QSDC) \cite{ping-pong,lm05}, and
deterministic secure quantum communication (DSQC) \cite{dsqc_summation,dsqcqithout-max-entanglement,dsqcwithteleporta,entanglement-swapping,reordering1,the:cao-and-song,
the:high-capacity-wstate,With-Anindita-pla}. For the purpose of the present study, all these schemes of secure
quantum communication can be broadly divided in two classes: Class
A: single-qubit-based schemes which do not use entangled states
to implement the protocol, like BB84 protocol \cite{bb84}, B92 protocol
\cite{b92}, LM05 protocol \cite{lm05}, and Class B: entangled-state-based
protocols, which uses one or more entangled states to implement the
protocol. Ekert protocol \cite{ekert}, BBM protocol \cite{BBM},
ping-pong (PP) protocol \cite{ping-pong},  are some of the protocols belonging to Class B. In
fact, there exists a one to one map between the protocols of Class
A and Class B. In principle, any task that can be implemented using
single qubit states can also be implemented using an entangled-state-based scheme. Of course, device independent schemes can be realized
only using the protocols of Class B. However, we  do not wish to stress
on that feature (device independence) here. Excluding ideas
of device independence, it can be shown that the security provided by
a scheme of Class A and the corresponding scheme of Class B is equivalent
in the ideal situation, where noise is not present. To illustrate
this point in Table \ref{tab:class1 and 2}, we have listed protocols
of Class A and Class B for various tasks related to secure quantum
communication. As we have already mentioned in an ideal situation,
these schemes (i.e., any two schemes shown in the same row of Table
\ref{tab:class1 and 2}) are equivalent as  far as the ability to perform
the cryptographic task in a secure manner is concerned. However, to
the best of our knowledge this equivalence is not investigated in
the realistic situation (i.e., in the presence of noise). Keeping
this fact in mind, this paper aims to perform a comparative study
of the protocols for secure quantum communication under various noise
models. Specifically, we wish to compare single-qubit-based protocols
(protocols of Class A) with entangled-state-based protocols (Protocol
of Class B) under various noise models. Here, it may be noted that
although, such comparative study has not yet been performed for protocols
of Class A and Class B mentioned above, a similar comparative study
has been performed on conjugate-coding-based protocols of secure quantum communication and orthogonal-state-based
protocols of secure quantum communication, which are equivalent in the ideal situation (\cite{CS-thesis} and references therein), but not in noisy
environment (\cite{decoy} and references therein). Further, there are various equivalent but different
decoy-qubit-based strategies (such as the BB84 subroutine, GV subroutine)
for eavesdropping checking that are used in standard protocols of
secure quantum communication. These subroutines are also known to
be equivalent in a noise free environment, but a recent study has
established that they are not equivalent in a noisy environment \cite{decoy}.
This recent observation has further motivated us to perform the present
investigation and to systematically investigate the effect of different
type of noises on various type of schemes of secure quantum communication.

\begin{table}
\begin{centering}
\begin{tabular}{|c|>{\centering}p{3cm}|>{\centering}p{4cm}|>{\centering}p{4cm}|}
\hline 
Sr. No. & Quantum Cryptographic Task & Protocol from Class A & Protocol from Class B\tabularnewline
\hline 
1 & QKD & B92 protocol \cite{b92} & BBM protocol \cite{BBM}\tabularnewline
\hline 
2 & QKA & Chong et al. protocol \cite{QKA-singlepar} & Shukla et al. protocol \cite{Qka-CS}\tabularnewline
\hline 
3 & QSDC & LM05 protocol \cite{lm05} & PP protocol \cite{ping-pong}\tabularnewline
\hline 
4 & QD & Shi et al. protocol \cite{QD-singlepar} & Ba An protocol \cite{ba-an}\tabularnewline
\hline 
\end{tabular}

\protect\caption{\label{tab:class1 and 2}Single-qubit-based and entangled-state-based
protocols for various tasks related to secure quantum communication.}
\par \end{centering}
\end{table}

There are several noise models \cite{nielsen,preskill}. Here, we
will restrict ourselves to the study of the effects of amplitude damping
(AD) channel, phase damping (PD) channel \cite{sbghosh, sbomkar}, collective noise and Pauli
noise. Finally, we will also discuss squeezed generalized amplitude
damping (SGAD) channel \cite{sbomkar,SGAD, sbsrikGP} and note that results for the generalized amplitude
damping (GAD) channel as well as that for the AD channel can be obtained from the results computed for the
SGAD channel. The motivation to study these noise models is that the
AD noise model deals with an interaction of the quantum system with
a zero temperature (vacuum) bath. An energy dissipation is involved
in this noise model while not in PD. These two noise models can bring about the phenomena of
entanglement decay and entanglement sudden-death \cite{ent-sud-death}.
Here, as we wish to analyze the equivalence between a single-qubit-based scheme with an entanglement-based one, these two noise models
become relevant. Collective noise is a coherent effect on all the
qubits, viz., all the polarization encoded photons traveling through
an optical fiber undergo the same birefringence \cite{col-exp}. The
Pauli noise channels include various physically relevant cases, such
as bit flip, phase flip, and depolarizing channels \cite{sbomkar,Pauli-exp,Pauli-ch-est,Pauli-ch-cap}.
SGAD channels are a generalization of the AD family of channels, which includes the GAD and involves the dissipative 
interaction with a non-zero temperature bath with non-vanishing
squeezing \cite{SGAD}. The squeezing, being a quantum resource, provides
an edge over GAD channels, which study a dissipative interaction with
a finite temperature bath without squeezing \cite{switch,our-QDs,tomogram}.
Hence, the choice of SGAD channel enables investigations into both non-zero
as well as vanishing regimes of squeezing. The wide applicability of all
these noise models sets our motivation to systematically study various
schemes for secure quantum communication under noisy environment and to
analyze their equivalence.

The remaining part of the paper is organized as follows. In Section
\ref{sec:noise-models}, we briefly discuss the noise models we are
going to apply on the schemes mentioned in Table \ref{tab:class1 and 2}.
The next section is dedicated to the method adopted to study the effect of noise
models described in Section \ref{sec:noise-models}. In Section \ref{sec:Various-protocols},
we briefly describe the protocols listed in Table \ref{tab:class1 and 2},
and report the effect of various type of noises on these protocols with
a clear aim to compare single-qubit-based scheme for a specific cryptographic
task with the corresponding entangled-state-based scheme. Finally,
we conclude in Section \ref{sec:Conclusion}.

\section{Different noise models \label{sec:noise-models}}

The most important and widely studied noise models are the AD, PD, collective and Pauli noise models. Apart from
these, generalization of AD considering
a dissipative interaction with a thermal and squeezed thermal bath
have been studied as GAD and SGAD, respectively. Here, we describe
only the SGAD channels as the effect of the GAD channel can
be obtained as its limiting case for zero bath squeezing. Further, as the AD noise is a limiting
case of GAD, it provides a consistency check of the obtained results
under SGAD noise. In what follows, we will study the effect of all these
noise models on the protocols of secure quantum communication that are listed in Table \ref{tab:class1 and 2}. The
noise models we have opted to study in the present paper are briefly
described below.

\subsection{AD noise model}

The AD noise simulates the dissipative interaction of a quantum system
with a vacuum bath. A perception about the importance
of this noise model can be obtained easily if we consider the large number
of theoretical and experimental works on this noise model reported
in the recent past (\cite{decoy,ent-sud-death,our-QDs,AD-prot-exp,exp-noise,exp-sr}
and references therein). The Kraus operators of an AD
channel are given by \cite{nielsen,preskill, SGAD} 
\begin{equation}
E_{0}=\left[\begin{array}{cc}
1 & 0\\
0 & \sqrt{1-\eta}
\end{array}\right],\,\,\,\,\,\,\,\,\,\,\,\,\,\,\,E_{1}=\left[\begin{array}{cc}
0 & \sqrt{\eta}\\
0 & 0
\end{array}\right],\label{eq:Krauss-amp-damping}
\end{equation}
where $\eta$ ($0\leq\eta\leq1$) is the probability of error or decoherence
rate.

\subsection{PD noise model}

Similarly, Kraus operators for phase-damping noise model are \cite{nielsen,preskill, sbomkar}
\begin{equation}
E_{0}=\sqrt{1-\eta}\left[\begin{array}{cc}
1 & 0\\
0 & 1
\end{array}\right],\,\,\,\,\,\,\,\,\,\,\,\,\,\,\,E_{1}=\sqrt{\eta}\left[\begin{array}{cc}
1 & 0\\
0 & 0
\end{array}\right],\,\,\,\,\,\,\,\,\,\,\,\,\,\,\,E_{2}=\sqrt{\eta}\left[\begin{array}{cc}
0 & 0\\
0 & 1
\end{array}\right],\label{eq:Krauss-phase-damping}
\end{equation}
where $\eta$ ($0\leq\eta\leq1$) is the decoherence rate. This is
another widely studied noise model. For instance, PD noise is discussed
in Refs. (\cite{decoy,sbomkar, sbsrikGP,ent-sud-death,exp-noise,exp-sr,CBRSP-our-paper,jay-cum,crypt-switch}
and references therein). This noise model is also experimentally simulated
in Refs. \cite{exp-noise,exp-sr}.

\subsection{Collective noises}

A coherent effect of environment on all the travel qubits passing
through a channel \cite{col-prl} 
can be studied using collective rotation (CR) and dephasing (CD) noise
models. It is known that the singlet states are resistant to an arbitrary
collective noise \cite{col-prl}. Recently, the effect of collective
noise on various schemes of quantum communication has been studied
\cite{QD-hwang,decoy,col-exp,col-adv,col-Bell,coll-noise}. Interestingly,
these studies provided protocols for quantum communication,
which use logical qubits to avoid the effect of collective noise (cf. \cite{QD-hwang,col-Bell,coll-noise}). Before we proceed further let us briefly introduce CR and CD noise models.

\subsubsection{CR noise model}

CR noise transforms $\left|0\right\rangle \rightarrow\cos\theta\left|0\right\rangle +\sin\theta\left|1\right\rangle $
and $\left|1\right\rangle \rightarrow-\sin\theta\left|0\right\rangle +\cos\theta\left|1\right\rangle $.
Here, $\theta$ is the noise parameter \cite{decoy,col-adv,col-Bell,coll-noise}.
Mathematically, a rotation operator acts on the quantum state of travel
qubits corresponding to this transformation.

\subsubsection{CD noise model}

CD noise leaves $\left|0\right\rangle $ unchanged while transforms
$\left|1\right\rangle $ as $\left|1\right\rangle \rightarrow\exp\left(i\phi\right)\left|1\right\rangle $,
where $\phi$ is the noise parameter \cite{decoy,col-adv,col-Bell,coll-noise}.
This is equivalent to a phase gate.

\subsection{Pauli noise}

The set of all Pauli channels is a tetrahedron. The phase flip and phase damping channels  correspond to a proper subset of the Pauli channels.
Depolarizing channels forms a 1-simplex embedded within the convex polytope representing the Pauli channels \cite{sbomkar}.
Pauli noise \cite{sbomkar} is studied using operators $E_{i}=\sqrt{p_{i}}\sigma_{i},$
where $\sigma_{0}=\mathbb{I}$, $\sigma_{1}=X$, $\sigma_{2}=iY$,
and $\sigma_{3}=Z$. Here, $p_{i}$ corresponds to the probability
with which a particular Pauli operation is applied \cite{Pauli-exp,Pauli-ch-est,Pauli-ch-cap}.
Corresponding expression for the depolarizing channel can be obtained
with $p_{i}=\frac{p^{\prime}}{3}$ for $i\in\left\{ 1,2,3\right\} $
and $p_{0}=1-p^{\prime}$. Specifically, it would mean that with a certain
probability the state remains unchanged while with the remaining probability, it becomes completely mixed. 
Further, information regarding bit flip, phase flip and bit-phase flip channels can be obtained with $p_{0}=1-p^{\prime}$
and $p_{i}=p^{\prime}$ for $i=1,3$ and 2, respectively. This kind
of noise channel is studied for noise estimation \cite{Pauli-exp},
channel characterization \cite{Pauli-ch-est} and error correction
\cite{Pauli-ch-cap}.

\subsection{SGAD noise model}

SGAD channel is a generalization of the AD and GAD channels and is characterized by the following Kraus
operators \cite{sbomkar,SGAD}  
\[
\begin{array}{lcl}
E_{0} & = & \sqrt{Q}\left[\begin{array}{cc}
1 & 0\\
0 & \sqrt{1-\lambda\left(t\right)}
\end{array}\right],\end{array}
\]
\[
\begin{array}{lcl}
E_{1} & = & \sqrt{Q}\left[\begin{array}{cc}
0 & \sqrt{\lambda\left(t\right)}\\
0 & 0
\end{array}\right],\end{array}
\]
\[
\begin{array}{lcl}
E_{2} & = & \sqrt{1-Q}\left[\begin{array}{cc}
\sqrt{1-\nu\left(t\right)} & 0\\
0 & \sqrt{1-\mu\left(t\right)}
\end{array}\right],\end{array}
\]
\begin{equation}
\begin{array}{lcl}
E_{3} & = & \sqrt{1-Q}\left[\begin{array}{cc}
0 & \sqrt{\mu\left(t\right)}e^{-i\Phi\left(t\right)}\\
\sqrt{\nu\left(t\right)} & 0
\end{array}\right],\end{array}\label{eq:SGAD-kraussoperators2-1}
\end{equation}
where $\lambda\left(t\right)=\frac{1}{p}\left\{ 1-\left(1-p\right)\left[\mu\left(t\right)+\nu\left(t\right)\right]-\exp\left(-\gamma_{0}\left(2N+1\right)t\right)\right\} ,$
$\mu\left(t\right)=\frac{2N+1}{2N\left(1-p\right)}\frac{\sinh^{2}\left(\gamma_{0}at/2\right)}{\sinh^{2}\left(\gamma_{0}\left(2N+1\right)t/2\right)}\exp\left(-\frac{\gamma_{0}}{2}\left(2N+1\right)t\right),$
and $\nu\left(t\right)=\frac{N}{\left(1-p\right)\left(2N+1\right)}\left\{ 1-\exp\left(-\gamma_{0}\left(2N+1\right)t\right)\right\} $.
Here, $\gamma_{0}$ is the spontaneous emission rate, $a=\sinh\left(2r\right)\left(2N_{th}+1\right),$
and $N=N_{th}\left\{ \cosh^{2}\left(r\right)+\sinh^{2}\left(r\right)\right\} +\sinh^{2}\left(r\right),$
where $N_{th}=1/\left\{ \exp\left(\hbar\omega/k_{B}T\right)-1\right\} $
 and $\Phi\left(t\right)$
is equal to the bath squeezing angle. The analytic expression for
the parameter $Q$  is quite involved, and can be obtained from Ref. \cite{SGAD}.
The beauty of SGAD channel is that for zero bath squeezing ($\Phi$), 
it reduces to GAD channel, which can
further be reduced to zero temperature bath (AD channel), where $Q$
becomes 1. Hereafter, we will avoid the time $t$ in the argument
of all the expressions under SGAD noise for simplicity of notations.
Quasiprobability distributions and tomogram of the single and two
qubit spin states under the SGAD channels have been studied recently
in \cite{our-QDs,tomogram}. The influence of SGAD noise on a quantum cryptographic switch
was analyzed in \cite{switch}.

\section{Strategy for studying the effect of various noise models on the protocols
of secure quantum communication \label{sec:Strategy-noise}}

The effect of noise can be studied by using a distance-based measure,
fidelity, between the final quantum state expected in the absence
of noise and the final state obtained when one of the noise models discussed
above is considered. To be precise, the strategy adopted in Ref.
\cite{decoy,CBRSP-our-paper,crypt-switch} will be used here. Before
we discuss various protocols of secure quantum communication and
the effect of noise on them, we will briefly summarize the
strategy adopted for the task. 

Consider an initial pure state $\rho=|\psi\rangle\langle\psi|$ which is to be
evolved under a noisy environment. The evolution of the state after
applying the Kraus operators characterizing a particular noise is $\rho_{k}=\underset{i}{\Sigma}E_{i}\rho E_{i}^{\dagger}$,
where $E_{i}$s are the Kraus operators for the chosen noise model under consideration.
Specifically, the Kraus operators of AD, PD, SGAD and Pauli channels
are given in Section \ref{sec:noise-models}. 

Further, in case of the coherent effect of noise on all the qubits,
i.e., collective noise, the transformed state is obtained as $\rho_{k}=U\rho U^{\dagger}$,
where $U$ is the unitary operation due to corresponding noise. The
unitary operations for both collective noises are given in the previous
section.

Finally, fidelity, defined as 
\[
F=\langle\psi|\rho_{k}|\psi\rangle,
\]
between the final state after the effect of noise $\rho_{k}$ and
pure initial state $|\psi\rangle$ is used as a measure of the effect
of noise. It would be worth mentioning here that the fidelity expression
used here has been used in Refs. (\cite{decoy,rsp-with-noise,fdly} and references therein). However, conventionally, an equivalent, but a slightly different  
definition of fidelity is used, and fidelity for two quantum states $\rho$ and $\sigma$ is defined as 
$F(\sigma,\rho)=Tr\sqrt{\sigma^{\frac{1}{2}}\rho\sigma^{\frac{1}{2}}}$. 

In the current study, we have assumed that one of the noise models
is studied at a time. Further, we have also considered that only the travel qubits
are affected by the environment, while the qubits not traveling through
the channel, i.e., home qubits remain unaffected.

\section{Various aspects (protocols) of secure quantum communication and effect
of noise on them \label{sec:Various-protocols}}

We briefly review two protocols for each type of
secure quantum communication task (namely, QKD, QKA, QSDC and QD),
and study the effect of the above described noise models on them. For this
we chose one protocol from Class A and
another one from Class B. Specifically, for a cryptographic task listed
in the second column of Table \ref{tab:class1 and 2}, a protocol
from Class A (B) is mentioned in the third (fourth) column. Here,
we aim to compare the protocol mentioned in the third column of Table
\ref{tab:class1 and 2} with the protocol mentioned in the fourth column
of the same row under different type of noise models. The purpose,
is to investigate their equivalence when subjected to different noise
models discussed in Section \ref{sec:noise-models}. Specifically,
the strategy mentioned in the previous section is used here to
perform the comparison by comparing fidelity. We obtain 
expressions of fidelity for the quantum
states to be recovered at the end of each protocol. Further, we would
like to mention that all the fidelity expressions reported here are
obtained as an average fidelity for all possible choices of initial states
and encoding on them. For example, if we consider Ba An protocol of QD where a predecided entangled state is used as initial state, then there will be 16 possible cases 
as Alice and Bob each can encode messages using 4 different operations. Similarly, for a single-qubit-based QD scheme there are 16 possible cases with 4 initial states and 2 possible encodings
by each party. This is why for each type of QD average fidelity is obtained by computing fidelity for all cases and then averaging. A similar approach is adopted in the rest of the paper to obtain average fidelity for various protocols.  

\subsection{QKD protocols and effect of noise on them}

Here, as we compare a single-qubit-based scheme for QKD with
a QKD scheme which requires an entangled state. Specifically, we opt
for B92 protocol \cite{b92} as an example of single-qubit-based QKD
scheme and BBM protocol \cite{BBM} as its entangled state counterpart.

\subsubsection{B92 protocol }

A modified version of BB84 with less resources was proposed by Bennett
in 1992 \cite{b92}. Hence, the protocol is referred to as B92 protocol.
The B92 scheme can be summarized in the following steps:
\begin{description}
\item [{B92~1:}] Alice sends a random string of $|0\rangle$ and $|+\rangle$
to Bob, where it is assumed that $|0\rangle$ and $|+\rangle$
correspond to bit values 0 and 1, respectively. \\
We can easily observe the modification
from BB84 as, in BB84 a random string of $\left\{ |0\rangle,|1\rangle,|+\rangle,|-\rangle\right\} $
was prepared. 
\item [{B92~2:}] Bob measures the received qubits in either computational
$\left\{ |0\rangle,|1\rangle\right\} $ or diagonal $\left\{ |+\rangle,|-\rangle\right\} $
basis randomly.\\
Here, Bob does not announce his choice of basis, which is in contrast to BB84.
\item [{B92~3:}] From his measurement outcome Bob keeps only the qubits
with measurement outcome $|1\rangle$ or $|-\rangle$ and announces
the same. Subsequently, Alice also discards the rest of the qubits.
The reason behind discarding the measurement outcomes can be understood by noting
that a contribution to measurement outputs $|0\rangle$ or $|+\rangle$
can be from both the initial states $|0\rangle$ and $|+\rangle$, due to which the measurement outcomes $|0\rangle$ or $|+\rangle$
can lead to a non-conclusive result. Therefore, only $|1\rangle$
or $|-\rangle$ outcomes are considered which correspond to Alice's
bit values 1 and 0. Hence, these qubits can be used to generate a
random symmetric key. 
\item [{B92~4:}] Bob announces the measurement outcomes of a part of the
generated string with the positions of the qubits for verification
of eavesdropping. For the corresponding qubits Alice checks the measurement
outcome with the initial state as $|0\rangle_{A}\rightarrow|-\rangle_{B}$
and $|+\rangle_{A}\rightarrow|1\rangle_{B}$. For the errors above a tolerable limit the protocol is discarded. Otherwise a secure
and symmetric key can be generated between the two users. 
\end{description}
The protocol described above can be studied under various noise models.
When the qubit prepared by Alice travels to Bob under the effect of
AD noise, the obtained fidelity is
\begin{equation}
F_{AD1}^{QKD}=\frac{1}{4}\left(\sqrt{1-\eta}+3\right).
\end{equation}
Here, and in the remaining part of the paper, required expressions
of fidelity are provided using a notation of the form $F_{ji}^{x}$,
where $j:j\in\{{\rm AD,\,PD\,,CR\,,CD\,,SGAD\,,P}\}$ is the type
of noise model; $i$ is 1 and 2 for single-qubit-based and entanglement-based schemes, respectively; and $x$ denotes the type of secure quantum
communication, i.e., $x\in\{{\rm QKD,\,QKA,\,QSDC,\,QD}\}$. Now,
considering that travel qubits have propagated via a PD channel, we
obtain

\begin{equation}
F_{PD1}^{QKD}=\frac{1}{4}\left(-\eta+4\right).
\end{equation}
In the collective noisy environment, the obtained fidelity expressions
are

\begin{equation}
F_{CD1}^{QKD}=\frac{1}{4}(\cos(\text{\ensuremath{\phi}})+3),
\end{equation}
and

\begin{equation}
F_{CR1}^{QKD}=\cos^{2}(\text{\ensuremath{\theta}}),
\end{equation}
for CD and CR noise channels, respectively. The analytic expressions
of fidelity under the effect of Pauli and SGAD channels are

\begin{equation}
F_{P1}^{QKD}=\frac{1}{2}(2p_{1}+p_{2}+p_{4})
\end{equation}
 and

\begin{equation}
F_{SGAD1}^{QKD}=\frac{1}{4}\left(\sqrt{1-\mu}\sqrt{1-\nu}-2\nu+Q\left(\sqrt{1-\lambda}-\sqrt{1-\mu}\sqrt{1-\nu}+2\nu\right)-\sqrt{\mu}\sqrt{\nu}(Q-1)\cos(\Phi)+3\right),
\end{equation}
respectively.

\subsubsection{BBM protocol }

\begin{figure}
\begin{centering}
\includegraphics[scale=0.75,angle=0]{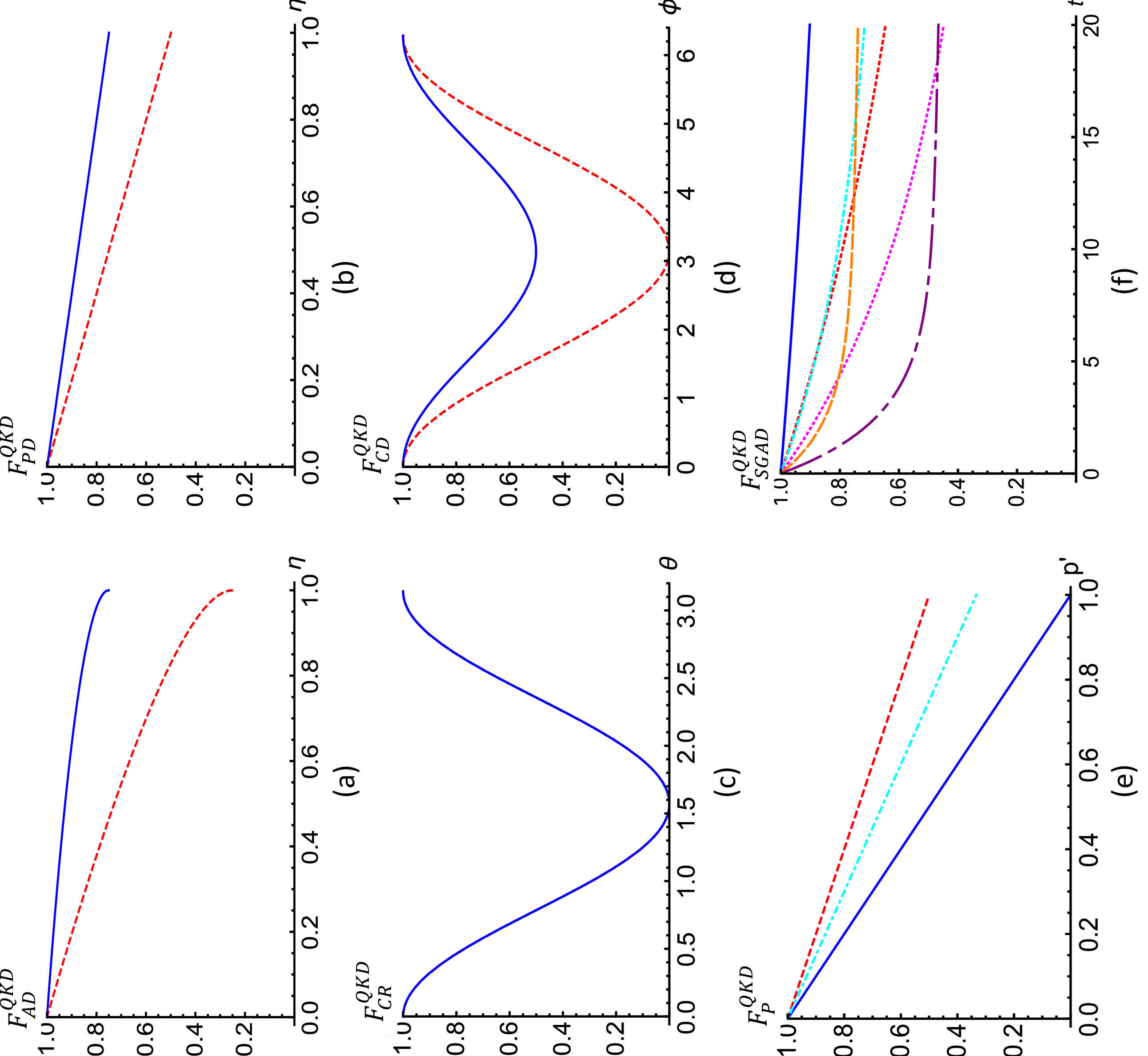}
\protect\caption{\label{fig:QKD-2D}QKD in AD, PD, CR and CD noises is shown in (a), (b), (c) and (d), respectively. The smooth (blue) and dashed (red) lines in all these plots correspond
to B92 and BBM protocols, respectively. For CR noise both B92 and BBM
protocols have the same fidelity (in (c)).  In (e),
the smooth (blue) line corresponds to the fidelity variation with
probability of bit-phase flip error for B92 protocol. This same curve
also illustrates the dependence of fidelity on probability in all four
possible cases discussed in the text for BBM protocol. The dashed
(red) and dotted dashed (cyan) lines show fidelity variation in B92
scheme with bit/phase flip error and depolarizing channel, respectively.
(f) demonstrates the effect of AD (in smooth (blue) and dashed
(red) lines); GAD (in dotted dashed (cyan) and dotted (magenta)
lines) with temperature $T=1$; and SGAD (in large dashed (orange) and large dotted dashed (purple) lines)
with $T=1$ and squeezing parameters $r=1$ and $\Phi=\frac{\pi}{8}$
for B92 and BBM protocols, respectively.}
\par \end{centering}
\end{figure}

The BBM protocol \cite{BBM} is a variant of the Ekert protocol \cite{ekert},
with reduced resources. Specifically, Ekert protocol uses three mutually
unbiased bases (MUBs) to calculate the correlation function for detecting
eavesdropping when the entanglement source was kept in between the
two authenticated users Alice and Bob \cite{ekert}. In contrast,
in BBM protocol, the source of entangled photon is given to Alice
and the requirement of three MUBs are reduced to two \cite{BBM}.
The protocol can be summarized as follows.
\begin{description}
\item [{BBM~1:}] Alice prepares a string of the singlet state $|\phi^{-}\rangle=\frac{|01\rangle-|10\rangle}{\sqrt{2}}$
and sends the second qubit to Bob keeping the first qubit with
herself. 
\item [{BBM~2:}] Both Alice and Bob measure their qubits of shared quantum
state in either computational $\left\{ |0\rangle,|1\rangle\right\} $
or diagonal $\left\{ |+\rangle,|-\rangle\right\} $ basis randomly.
Both the users announce their choices of measurement basis, but not
the measurement outcomes. 
\item [{BBM~3:}] Both users decide to discard the measurement outcomes
where their choices of measurement basis were different, as in all
the remaining cases their measurement outcomes are supposed to be
correlated.
\item [{BBM~4:}] Finally, both the users choose around half of the string
of the undiscarded instances and announce corresponding measurement
outcomes. If the error in the measurement outcomes is below certain tolerable limit both Alice and
Bob can obtain a symmetric key using the outcomes of the measurements performed on the remaining qubits which are not used for eavesdropping check. In other words, a lack of correlation in the measurement
outcomes is a signature of the presence of an adversary.
\end{description}
The entanglement-based protocol of QKD considered here, i.e., BBM
scheme, under the AD, PD, CD and CR noises lead to the following fidelities
expressions 
\begin{equation}
F_{AD2}^{QKD}=\frac{1}{4}\left(-\eta+2\sqrt{1-\eta}+2\right),
\end{equation}

\begin{equation}
F_{PD2}^{QKD}=\frac{1}{2}\left(-\eta+2\right),
\end{equation}

\begin{equation}
F_{CD2}^{QKD}=\cos^{2}\left(\frac{\phi}{2}\right),
\end{equation}
and

\begin{equation}
F_{CR2}^{QKD}=\cos^{2}(\theta),
\end{equation}
respectively. In case of Pauli channels, the fidelity only depends
on the probability with which the state remains unchanged

\begin{equation}
F_{P2}^{QKD}=p_{1}.
\end{equation}
Hence, a linear plot is expected. When the qubits travel under the
dissipative SGAD channel, the compact form of fidelity is

\begin{equation}
F_{SGAD2}^{QKD}=\frac{1}{4}\left(2\sqrt{1-\mu}\sqrt{1-\nu}-\mu-\nu+Q\left(-\lambda+2\sqrt{1-\lambda}-2\sqrt{1-\mu}\sqrt{1-\nu}+\mu+\nu\right)+2\right).
\end{equation}
Now, we will try to make a comparative analysis of the obtained fidelities
expressions in the QKD protocols from the two classes. When the travel
qubits are subjected to AD noise we observe that B92 protocol performs
better than BBM protocol for all values of decoherence rate $\eta$.
Quite similar nature is observed with PD noise as well. The gradual
decrease in the fidelity with the increasing decoherence rate for all these
cases can be observed in Fig. \ref{fig:QKD-2D} a and b. The fact
observed here is consistent with some of our recent observations that
single qubits perform better while traveling through AD and PD channels
\cite{decoy}. Interestingly, fidelity obtained for both the QKD schemes
considered here is the same when the travel qubits are subjected to
CR noise, as can be seen in Fig. \ref{fig:QKD-2D} c. In the presence of another type of collective
noise (namely CD noise), which is dephasing in nature, B92 is again seen to perform
better (cf. Fig. \ref{fig:QKD-2D} d). Here, it is worth noting that
the singlet state is decoherence free in an arbitrary collective noise when both the qubits of the single state travel through the noisy channel.
However, when one of the qubits travels through the channel having collective noise, it gets affected by the noise and as a consequence singlet state also gets affected by the noise. This is what we have observed here.

Expressions of fidelity under Pauli noise reveal that for BBM protocol
equally affected states will be obtained for bit-flip, phase-flip,
bit-phase flip errors with certain probability. For the
depolarizing channel the fidelity expression shows a similar nature.
The same expression is obtained for B92 protocol with bit-phase flip
error. Further, bit-flip and phase flip errors with equal probabilities
affect the state in a similar manner as the expression becomes same in both
the cases. The values are always higher than all others for the same amount
of error as shown in Fig. \ref{fig:QKD-2D} e. The obtained fidelity
in depolarizing channels can be seen to be intermediate between the
last two. 

From the fidelity expressions of SGAD noise, the corresponding fidelities under AD and GAD channels can be 
obtained as limiting cases. In Fig. \ref{fig:QKD-2D} f we
can see the advantage obtained due to squeezing. Specifically, for
non-zero squeezing we can obtain higher fidelity than that with a GAD
noisy environment for a longer time period. Further, it can also be observed
that a state is more affected while traveling through finite temperature
bath than in the vacuum bath (AD).

\subsection{QKA protocols and effect of noise on them}

In realistic scenarios, it may be preferable that a single party does
not control the whole key. In such scenarios QKD can be circumvented
by a key agreement protocol, where all the parties can equally contribute
in the final key. To be precise, QKA schemes are studied under two notions:
weaker and stronger. In the weaker notion of QKA protocols, the final
key is generated after negotiation between both the parties. If we follow this notion, then many of the
QKD schemes can be viewed as QKA schemes, such as BB84,
B92 and BBM discussed in the previous subsection. However, in the
strong notion all the parties contribute equally to the final shared
key. Many QKA schemes have been proposed in the
past (\cite{QKA,Qka-CS,QKA-singlepar} and references therein).

\subsubsection{Single-qubit-based QKA protocol }

A single-qubit-based quantum key agreement protocol given by Chong
et al., in 2010 \cite{QKA-singlepar} can be described in the following
steps:
\begin{description}
\item [{QKA1-1}] Alice randomly prepares an $n$ bit raw key $K_{A}$ and
a random string of 0 and 1.
\item [{QKA1-2}] Alice prepares $n$ qubits in such a way that for every 0
(1) in the key she prepares either $|0\rangle$ or $|+\rangle$ ($|1\rangle$
or $|-\rangle$) depending upon the corresponding bit value in the
random string 0 or 1, respectively. Finally, she sends all the qubits
to Bob.
\item [{QKA1-3}] Bob also prepares an $n$ bit raw key $K_{B}$. Now, to encode
this key he applies $I$ ($iY$) on the received qubits for 0 (1). 
\item [{QKA1-4}] Bob selects a random sequence from the qubits as verification
string and announces the positions of the corresponding qubits. He
also announces his raw key. 
\item [{QKA1-5}] Alice can extract a final key as $K=K_{A}\oplus K_{B}$
from her and Bob's keys. Subsequently, she broadcasts the obtained
values corresponding to the qubits Bob had chosen as verification
string along with the information of basis chosen for each qubit in
QKA1-2.
\item [{QKA1-6}] Using the information of the basis chosen Bob can also
extract the final key $K$. If the obtained values for Alice and Bob
have errors below a tolerable limit they share an unconditionally
secure quantum key. 
\end{description}
If the single-qubit-based QKA scheme described above is implemented using a quantum channel having AD noise then we obtain
\begin{equation}
F_{AD1}^{QKA}=\frac{1}{4}\left(-\eta+\sqrt{1-\eta}+3\right),
\end{equation}
whereas under PD noise we have

\begin{equation}
F_{PD1}^{QKA}=\frac{1}{4}\left(\eta^{2}-2\eta+4\right).
\end{equation}
On the effect of CD noise the fidelity becomes 

\begin{equation}
F_{CD1}^{QKA}=\frac{1}{4}(\cos(\text{\ensuremath{\phi}}_{1})+3),
\end{equation}
while under the influence of CR noise it is

\begin{equation}
F_{CR1}^{QKA}=\cos^{2}(\text{\ensuremath{\theta}}_{1}).
\end{equation}
In case the travel particles go through a Pauli channel the obtained
fidelity is

\begin{equation}
F_{P1}^{QKA}=\frac{1}{2}(2p_{1}+p_{2}+p_{4}).
\end{equation}
For an interaction with a squeezed thermal bath, the fidelity depends
on various parameters as 

\begin{equation}
F_{SGAD1}^{QKA}=\frac{1}{4}\left(\sqrt{1-\mu}\sqrt{1-\nu}-\mu-\nu+Q\left(-\lambda+\sqrt{1-\lambda}-\sqrt{1-\mu}\sqrt{1-\nu}+\mu+\nu\right)-\sqrt{\mu}\sqrt{\nu}(Q-1)
\cos(\Phi)+3\right).
\end{equation}

\subsubsection{Entangled-state-based QKA protocol }

There are various protocols of quantum key agreement that exploit
entanglement. Here, we wish to summarize
a protocol proposed by Shukla et al. in 2014 \cite{Qka-CS}. 
\begin{description}
\item [{QKA2-1}] Alice prepares $|\psi^{+}\rangle^{\otimes n}$,
where $|\psi^{+}\rangle\equiv\frac{1}{\sqrt{2}}(|00\rangle+|11\rangle)$.
She also prepares a raw key $K_{A}$ of $n$ bits.  She prepares a
string of all the first particles to be sent to Bob keeping all
the second qubits with herself. 
\item [{QKA2-2}] Alice prepares $\frac{n}{2}$ Bell states $|\psi^{+}\rangle^{\otimes\frac{n}{2}}$
as decoy qubits and concatenates them with the string of the first
particles of the Bell states and sends the $2n$ qubits to Bob after
applying a permutation operator $\Pi_{2n}$. 
\item [{QKA2-3}] After an authentic acknowledgment of the receipt of all
the qubits Alice announces the positions of the decoy qubits, i.e.,
information of $\Pi_{n}$. Using this information Bob performs a Bell
state measurement on partner pairs and calculates error rate. It
would be relevant to mention that the decoy-qubit-based security achieved
here with GV subroutine can be equivalently done by BB84 subroutine
where single qubit decoy qubits are used. They decide to proceed
if  error rates are below a certain value. 
\item [{QKA2-4}] Bob also prepares a raw key $K_{B}$. Further, on the
remaining qubits Bob encodes his raw key by applying $I$ or $X$
operations for 0 and 1, respectively. Subsequently, he prepares $\frac{n}{2}$
Bell states as decoy qubits and permutes the string of $2n$ qubits
by permutation operator $\Pi^{\prime}_{2n}$ after concatenating the decoy
and encoded qubits. Finally, he sends them to Alice.
\item [{QKA2-5}] Bob informs the coordinates of the decoy qubits using
which Alice computes the error rate. From this they choose whether
to proceed or not.
\item [{QKA2-6}] Alice announces her key publicly from which Bob can generate
the final key $K=K_{A}\oplus K_{B}$. 
\item [{QKA2-7}] Bob announces the permutation operator to rearrange the
particles in the encoded string with Alice. Using this Alice performs
a Bell state measurement on the partner pairs of home and travel qubits.
The measurement outcome would reveal Bob's key to Alice.
\item [{QKA2-8}] Alice can also obtain the final shared, unconditionally
secure, quantum key $K$.
\end{description}
The fidelity expression for the entanglement-based QKA scheme when
subjected to AD and PD noise are
\begin{equation}
F_{AD2}^{QKA}=\frac{1}{4}(\eta-2)^{2}
\end{equation}
and
\begin{equation}
F_{PD2}^{QKA}=\frac{1}{2}\left(\eta^{2}-2\eta+2\right),
\end{equation}
respectively. The quantum state evolves under the collective noise
such that the obtained fidelity with the expected pure state is

\begin{equation}
F_{CD2}^{QKA}=\frac{1}{2}\left\{ \cos(\text{\ensuremath{\phi}}_{1})\cos(\text{\ensuremath{\phi}}_{2})+1\right\}, 
\end{equation}
and

\begin{equation}
F_{CR2}^{QKA}=\frac{1}{2}\left\{ \cos^{2}(\text{\ensuremath{\theta}}_{1}-\text{\ensuremath{\theta}}_{2})+\cos^{2}(\text{\ensuremath{\theta}}_{1}+
\text{\ensuremath{\theta}}_{2})\right\}, 
\end{equation}
for CD and CR noise, respectively. Here, it may be noted that the
two noise parameters $\phi_{i}$ and $\theta_{i}$
correspond to each round of the travel qubit. The Pauli channels have
a symmetric expression for fidelity, given by

\begin{equation}
F_{P2}^{QKA}=p_{1}^{2}+p_{2}^{2}+p_{3}^{2}+p_{4}^{2}.
\end{equation}
The closed form analytic expression of fidelity, under the SGAD channel, for the above described QKA scheme \cite{Qka-CS} is

\begin{equation}
\begin{array}{lcl}
F_{SGAD2}^{QKA} & = & \frac{1}{4}\left\{ Q^{2}\left(\lambda^{2}-2\lambda(\mu+\nu+1)-2\left(2\sqrt{1-\lambda}\sqrt{1-\mu}\sqrt{1-\nu}+\mu+\nu-2\right)+\mu^{2}+5\mu\nu+\nu^{2}\right)\right.\\
 & + & \mu^{2}+\mu(5\nu-4)+(\nu-2)^{2}+\mu\nu(Q-1)^{2}\cos(2\Phi)\\
 & + & \left.2Q\left(\lambda(\mu+\nu-1)+2\sqrt{1-\lambda}\sqrt{1-\mu}\sqrt{1-\nu}-\mu^{2}+\mu(3-5\nu)-(\nu-3)\nu-2\right)\right\} .
\end{array}
\end{equation}

\begin{figure}
\begin{centering}
\includegraphics[scale=0.75,angle=0]{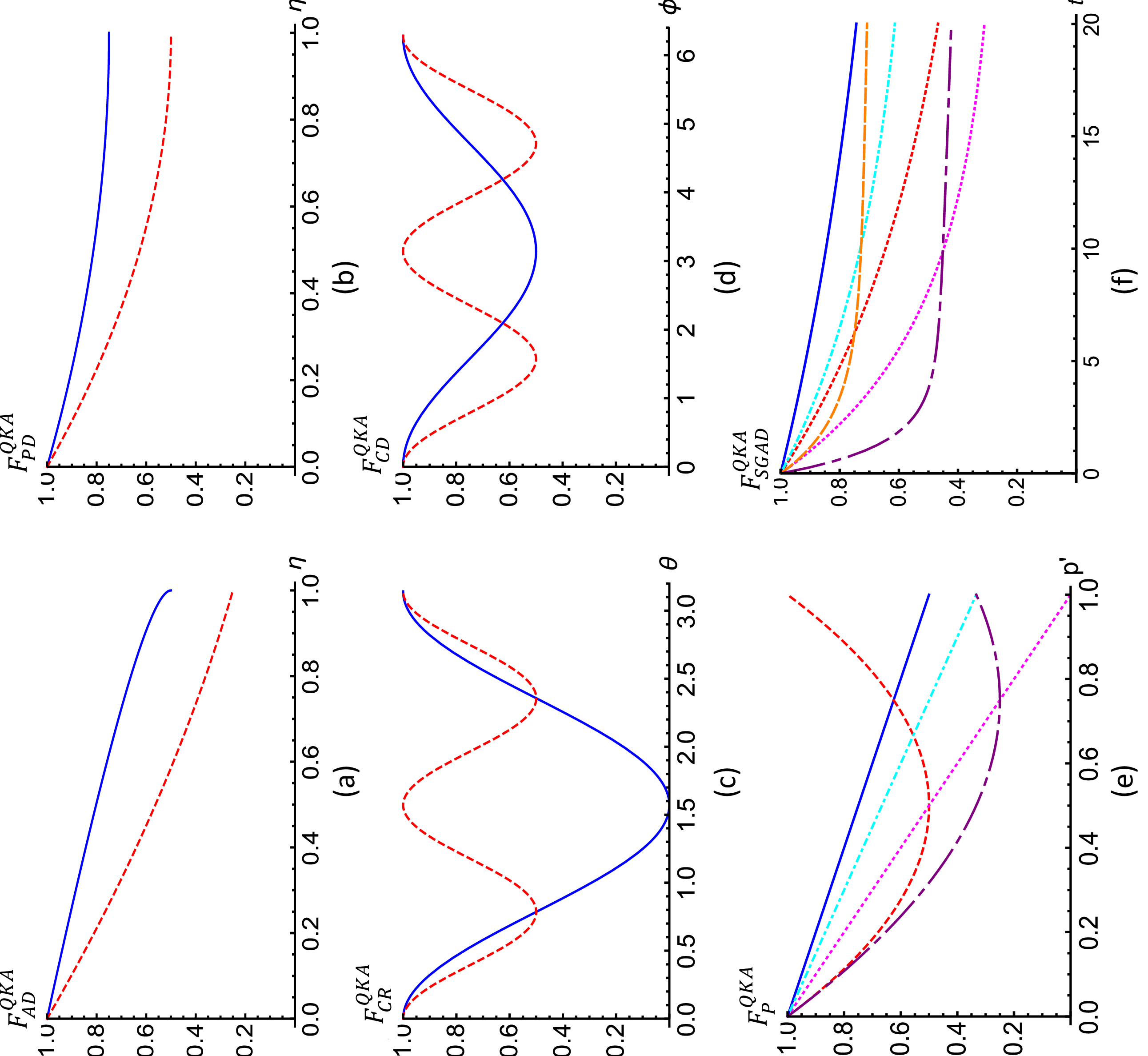}

\protect\caption{\label{fig:QKA-2D}(a)-(d) illustrates the fidelity obtained for QKA protocols when subjected to AD, PD, CR and CD noises,
respectively. The smooth (blue) and dashed (red) lines correspond
to single-qubit-based and entangled-state-based QKA protocols,
respectively. For CR and CD noises it is assumed that the noise parameter
is same for both the directions of travel of the qubit (i.e., Alice
to Bob and Bob to Alice). In (e), the effect of bit flip
error on single-qubit-based QKA and entangled-state-based QKA protocols are shown using smooth (blue)
and dashed (red) lines, respectively. In the same plot, dotted dashed (cyan) and large dotted dashed (purple)
lines correspond to the effect of depolarizing channel on Shi et al.'s and Shukla et al.'s QKA schemes, respecetively; and the dotted (magenta) line illustrates the effect of bit-phase flip error on the  single-qubit-based QKA scheme. (f) corresponds to
the effect of AD in smooth (blue) and dashed (red) lines; GAD 
in dotted dashed (cyan) and dotted (magenta) lines with $T=1$; and
SGAD in large dashed (orange)
and large dotted dashed (purple) lines with $T=1$ and squeezing parameters
$r=1$ and $\Phi=\frac{\pi}{8}$ for single-qubit-based and entangled-state-based QKA protocols, respectively.}
\par \end{centering}
\end{figure}

For QKA schemes two way quantum communication is involved unlike QKD
protocols in the previous section, where only sender to receiver communication
is involved. In both single-qubit-based and entangled-state-based QKA
schemes fidelity falls gradually with an increase in decoherence rate
$\eta$ when subjected to AD and PD noisy environments of identical strength (cf.
Fig. \ref{fig:QKA-2D} a and b). Similar to the QKD scheme,
single-qubit-based schemes perform better than the entangled-state-based ones in both these noisy channels. Further, this similarity
between the single-qubit-based QKD and QKA schemes for collective
noises, is depicted in the corresponding curves shown in Figs. \ref{fig:QKD-2D} c and d and Figs. \ref{fig:QKA-2D}
c and d. However, the entangled-state-based QKA scheme is seen to benefit under 
collective noise as fidelity for the entangled-state-based QKA scheme is 
more than that of single-qubit-based protocol, under the assumption
of the same noise strength in both rounds of the travel qubit. This fact
can be attributed to different choices of Bell state in entanglement-based QKD and QKA protocols.

\begin{figure}
\begin{centering}
\includegraphics[angle=270]{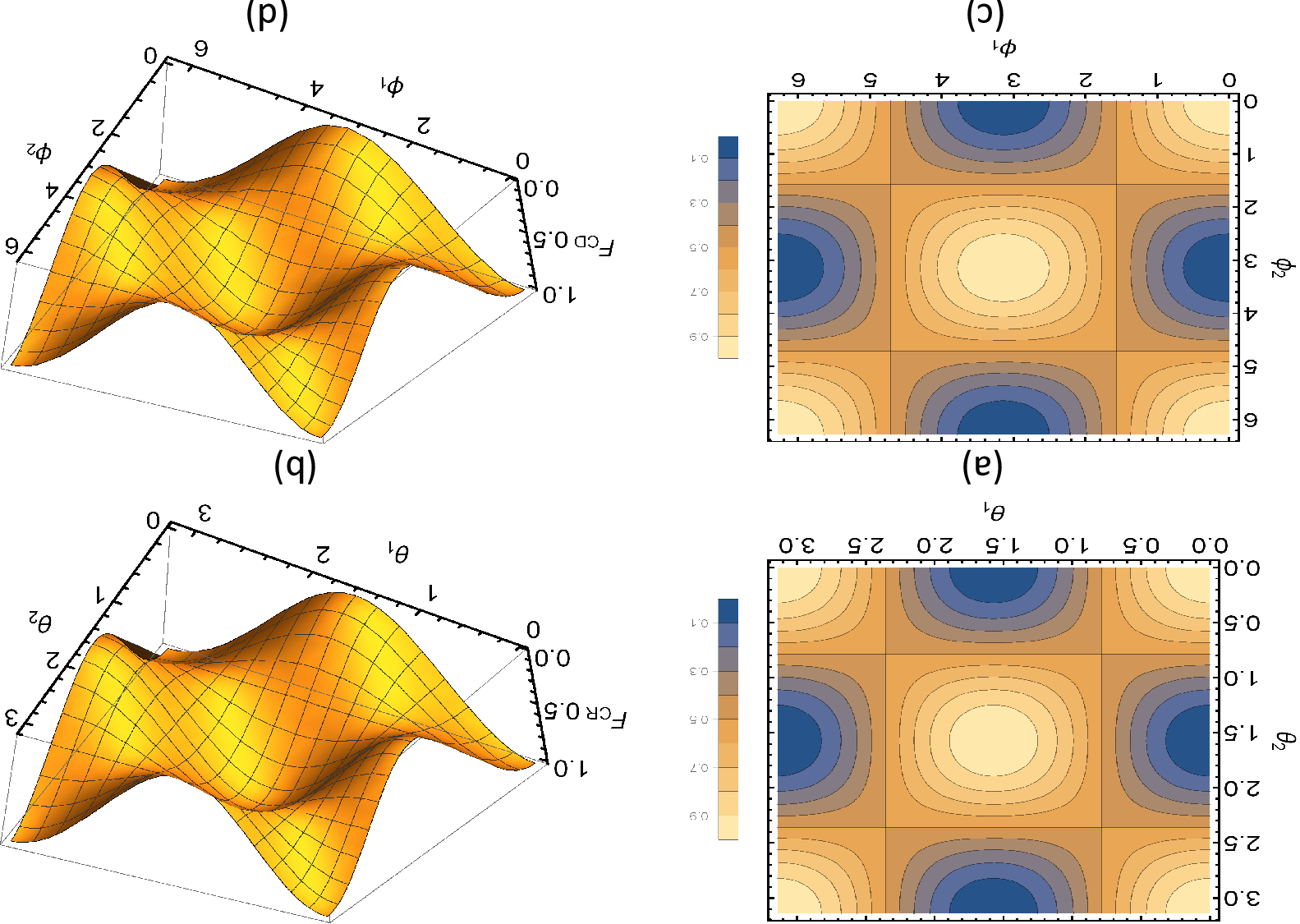}

\protect\caption{\label{fig:QKA-3D}Contour and 3D variation of fidelity of entangled-state-based QKA protocol is shown in CR and CD noises. Specifically, in (a) and (c) the contour plots under the effect of CR and CD noise are shown. Corresponding 3D plots can be seen in (b) and (d), respectively. The same
plots are obtained for PP protocol as well. A detailed discussion
follows in Subsection \ref{sub:QSDC-protocols}.}
\par \end{centering}
\end{figure}

Further, as discussed in the previous section, the collective noise
parameter remains the same for all the qubits traveling through a channel
at a particular time, but can have a different value at any other time.
The effect of two different values of noise parameters, of the collective
noises, on the fidelity of the obtained state can be studied by showing
either 3 dimensional variation or contour plots. Fig. \ref{fig:QKA-3D}
a and b (c and d) show both these kinds of plots for QKA scheme subjected
to CR (CD) noise. Hereafter, we will stick to the contour plots to
illustrate the effect of two parameters. Interestingly, it can be observed that
it is possible to obtain states with unit or null fidelity for some
values of noise parameters.

For the single-qubit-based scheme, the analytic expressions for fidelities are the same for all three types of Pauli channels (i.e., for bit flip, phase flip and 
bit-phase flip channels). The expressions of fidelity for bit flip and phase flip channels are also the same for  the entangled-state-based protocol, but for bit-phase 
flip error, we obtain a different expression for fidelity, and it is observed that the obtained value of fidelity is  smaller compared to the corresponding values for bit flip and phase flip errors. 
Fig. \ref{fig:QKA-2D} e shows
variations of fidelity in all these error channels, where an increase
in fidelity for entanglement-based QKA schemes can be attributed to
the presence of quadratic terms in the fidelity expression. 
The variation of fidelity in Fig. \ref{fig:QKA-2D}
f considering a dissipative interaction
via SGAD channel for both kinds of QKA schemes reemphasize the facts
established by their QKD counterparts (cf. Fig. \ref{fig:QKD-2D}
f). Specifically, with increase in temperature, dissipation increases,
causing decay in the fidelity of the recovered state. Also, squeezing turns out to be a
useful resource here, as observed from the increased fidelity of the SGAD plots compared to their GAD (without squeezing)
counterparts after a certian evolution period.

\subsection{QSDC protocols and effect of noise on them \label{sub:QSDC-protocols}}

Quantum secure communication not necessarily involves a key generation
or key agreement. There are direct communication protocols avoiding
key generation and such protocols are referred to as the protocols for secure direct quantum communication. These protocols can be broadly
categorized as QSDC and DSQC protocols depending upon the requirement
of additional classical communication for decoding of the information.
QSDC protocols do not require any additional classical communication
other than that involved in eavesdropping checking, while DSQC protocols
do. Here, we wish to discuss two QSDC protocols and compare them in the presence of noise.

\subsubsection{LM05 protocol}

A QSDC protocol without using entanglement was proposed by Lucamarini
and Mancini in 2005 which is now known as LM05 protocol \cite{lm05}. The
protocol can be briefly describe in the following steps:
\begin{description}
\item [{LM1}] Bob (receiver) prepares a random string of $|0\rangle,|1\rangle,|+\rangle,|-\rangle$
and sends it to Alice. 
\item [{LM2}] Alice chooses randomly half of the received qubits as a verification
string (to be used as decoy qubits) and performs eavesdropping checking
on these qubits. Specifically, Alice measures all the qubits in the
verification string in MUBs $\left\{ 0,1\right\} $ or $\left\{ +,-\right\} $
randomly. Then she announces the choice of basis with the position
of qubits. Based on this, Bob announces the qubits where he has chosen
the same basis to prepare the initial state. Depending on this, the
measurement outcomes of Alice are expected to be the same with the state prepared by
Bob in the absence of any attempt of eavesdropping. For errors below a tolerable
limit they proceed to the next step, else they start afresh.
\item [{LM3}] To encode her message on  half of the remaining qubits
Alice applies operator $I$ ($iY$) for sending 0 (1). Subsequently,
she returns the encoded qubits to Bob. Here, it would be nice to mention
that using such a scheme, for both choices of encoding, a particular
initial state will transform into orthogonal states. Consequently,
at Bobs end, a message can be easily decoded by measuring the state
in the basis it was prepared.
\item [{LM4}] Alice announces the coordinates of the qubits she had not
encoded on (as she wished to use them as decoy qubits for Alice to
Bob communication). Bob measures corresponding qubits in the basis
he had prepared them initially to check the presence of Eve for Alice
to Bob travel of the encoded particles. The same task can also be
achieved by Alice encoding on all the remaining qubits after eavesdropping
in LM2, while she prepares additional string of equal number of qubits
in $|0\rangle,|1\rangle,|+\rangle,|-\rangle$ randomly for eavesdropping
checking in this step.
\item [{LM5}] Only if Bob is convinced of the absence of Eve, he decodes
the message sent by Alice by measuring the qubits in the same basis
he had prepared them in LM1, otherwise they abort the protocol.
\end{description}
The effect of the AD noise on the single-qubit-based QSDC protocol (LM05 protocol)
opted here, LM05, can be deduced from the fidelity expression 
\begin{equation}
F_{AD1}^{QSDC}=\frac{1}{4}\left(\eta^{2}-3\eta+4\right).
\end{equation}
The corresponding expression under the effect of PD noise is

\begin{equation}
F_{PD1}^{QSDC}=\frac{1}{4}\left(\eta^{2}-2\eta+4\right).
\end{equation}
Similar to the entanglement-based QKA scheme, two rounds of quantum
communication is involved here, due to which the expressions of fidelity
under CD noise

\begin{equation}
F_{CD1}^{QSDC}=\frac{1}{4}(\cos(\text{\ensuremath{\phi}}_{1})\cos(\text{\ensuremath{\phi}}_{2})+3),
\end{equation}
and that for CR noise

\begin{equation}
F_{CR1}^{QSDC}=\cos^{2}(\text{\ensuremath{\theta}}_{1}+\text{\ensuremath{\theta}}_{2}),
\end{equation}
involve two noise parameters ($\phi_{1},\,\phi_{2}$ or $\theta_{1},\,\theta_{2}$) each. As usual, the fidelity expression
for Pauli channels with four parameters is

\begin{equation}
F_{P1}^{QSDC}=p_{1}^{2}+p_{2}^{2}+p_{3}^{2}+p_{4}^{2}+(p_{1}+p_{3})(p_{2}+p_{4}).
\end{equation}
The presence of quadratic terms is signature of two rounds of quantum
communication. When the travel qubits undergo a dissipative interaction
characterized by the SGAD channel, the fidelity is obtained
as

\begin{equation}
\begin{array}{lcl}
F_{SGAD1}^{QSDC} & = & \frac{1}{8}\left\{ 2\left((\nu-3)\nu+Q^{2}\left(-2\lambda\nu+(\lambda-1)\lambda+\nu^{2}-\nu+2\right)+2(\nu-1)Q(\lambda-\nu+1)+4\right)\right.\\
 & - & 4\sqrt{1-\lambda}\sqrt{1-\mu}\sqrt{1-\nu}Q^{2}+\mu(Q-1)(-7\nu+Q(-4\lambda+7\nu-2)\\
 & + & 4\sqrt{\mu}\sqrt{\nu}(Q-1)\cos(\Phi)\left(-\sqrt{1-\mu}\sqrt{1-\nu}-\sqrt{1-\lambda}Q+\sqrt{1-\mu}\sqrt{1-\nu}Q\right)\\
 & + & \left.\nu(Q-1)\cos(2\Phi)+6)+4\sqrt{1-\lambda}\sqrt{1-\mu}\sqrt{1-\nu}Q+2\mu^{2}(Q-1)^{2}\right\}.
\end{array}
\end{equation}

\subsubsection{Ping-pong protocol}

An entangled-state-based QSDC protocol was proposed by ${\rm Bostr\ddot{o}m}$
and Felbinger in 2002 \cite{ping-pong}. Precisely, LM05 protocol
is a single-qubit-based counterpart of PP protocol. The PP
protocol works as follows:
\begin{description}
\item [{PP1}] Bob prepares $|\psi^{+}\rangle^{\otimes n}$,
where $|\psi^{+}\rangle\equiv\frac{1}{\sqrt{2}}(|00\rangle+|11\rangle)$.
Then he sends all the first particles to Alice keeping all the second
qubits with himself. 
\item [{PP2}] Alice forms a verification string by randomly choosing a
set of $\frac{n}{2}$ qubits to perform BB84 subroutine as was done in LM2.
Specifically, Alice measures the qubits randomly in $\left\{ 0,1\right\} $
or $\left\{ +,-\right\} $ basis and announces the choice of basis.
Bob also measures his qubits in the same basis. In the absence of Eve,
their measurement outcomes are expected to be correlated. In the absence
of such a correlation they discard the protocol and return to PP1,
otherwise they proceed.
\item [{PP3}] Out of  half of the remaining qubits Alice randomly makes
two sets of equal number of qubits. One set for encoding her message and another set for
eavesdropping check for Alice to Bob communication. To encode 1
Alice applies $X$ gate before sending the qubit to Bob, and for sending
0 she returns the qubit unchanged.
\item [{PP4}] Alice informs the coordinates of verification string and
Bob performs BB84 subroutine to compute the error rate.
\item [{PP5}] For low error rates, Bob performs Bell state-measurement
on the partner pairs to decode the message sent by Alice.
\end{description}
The analytical expressions of fidelity in the case of the PP protocol exactly match those   
for entanglement-based QKA scheme \cite{Qka-CS}. Therefore, we avoid repetition of the expressions and
carry on with the discussion regarding the comparison between LM05 and
PP protocols under noisy environments.

\begin{figure}
\begin{centering}
\includegraphics[scale=0.75,angle=00]{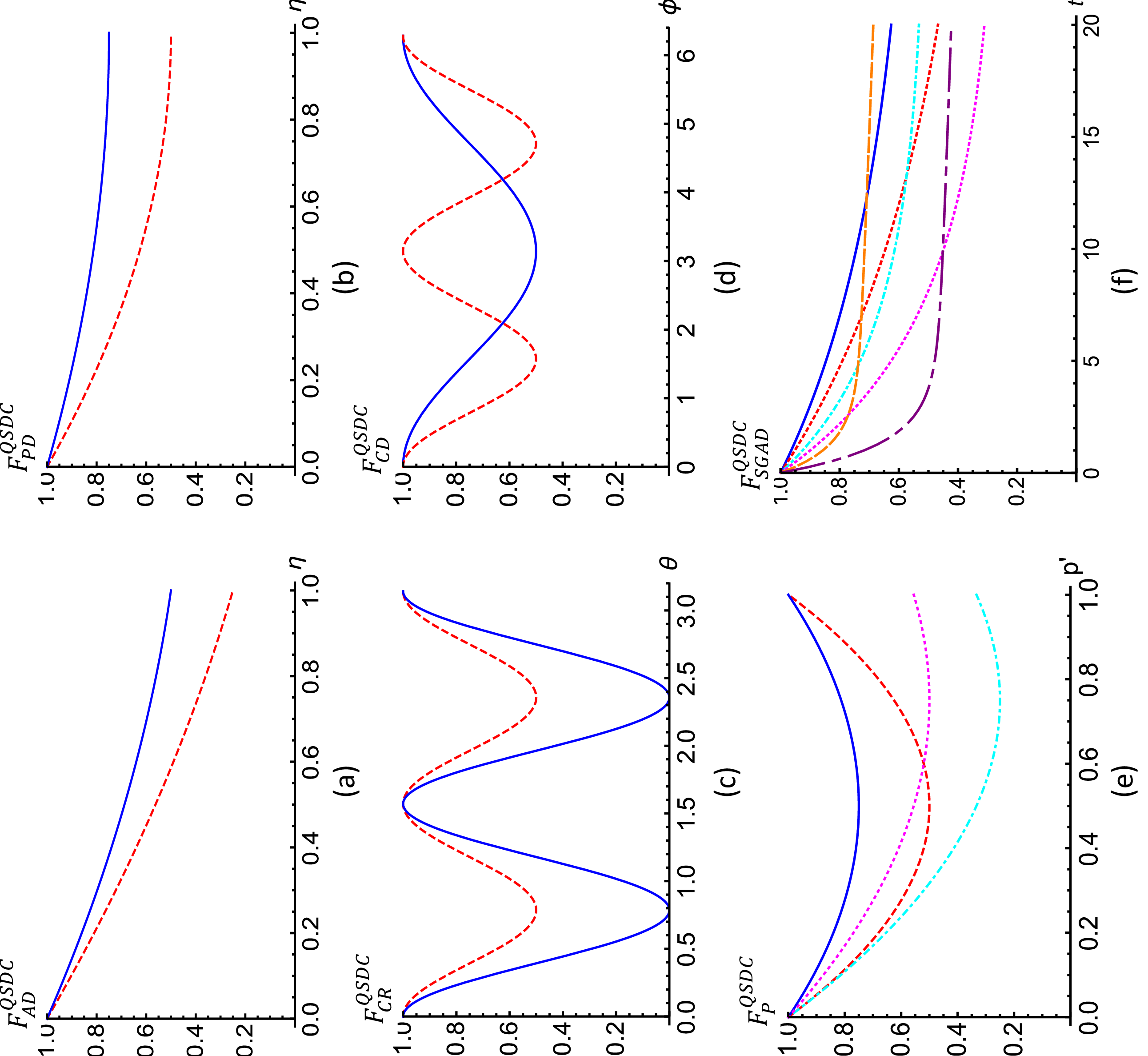}
\par\end{centering}

\protect\caption{\label{fig:QSDC-2D}QSDC under AD, PD, CR and CD noises are depicted in (a), (b), (c) and (d), respectively. The smooth (blue) and dashed (red) lines
correspond to LM05 and PP protocols, respectively. For CR and CD noises
it is assumed that the noise parameter is same for both the directions
of travel of the qubit (i.e., Alice to Bob and Bob to Alice). In (e), the fidelity under the depolarizing channel for LM05 and PP
protocols are shown in dotted (magenta) and dotted dashed (cyan) lines,
respectively. In all the remaining cases of PP and bit phase flip
for LM05 the red line illustrates the fall and revival in fidelity.
Lastly, the blue line corresponds to the fidelity in bit flip and
phase flip errors in LM05 scheme. (f) illustrates the
effect of AD (i.e., an interaction with a zero temperature and squeezing
bath) in smooth (blue) and dashed (red) lines; GAD (i.e., an interaction
with a non-zero temperature and zero squeezing bath) in dotted dashed
(cyan) and dotted (magenta) lines with $T=1$; and SGAD (finite temperature
and squeezing bath) in large dashed (orange) and large dotted dashed
(purple) lines with $T=1$ and squeezing parameters $r=1$ and $\Phi=\frac{\pi}{8}$
for LM05 and PP protocols, respectively.}
\end{figure}

\begin{figure}
\begin{centering}
\includegraphics[angle=270]{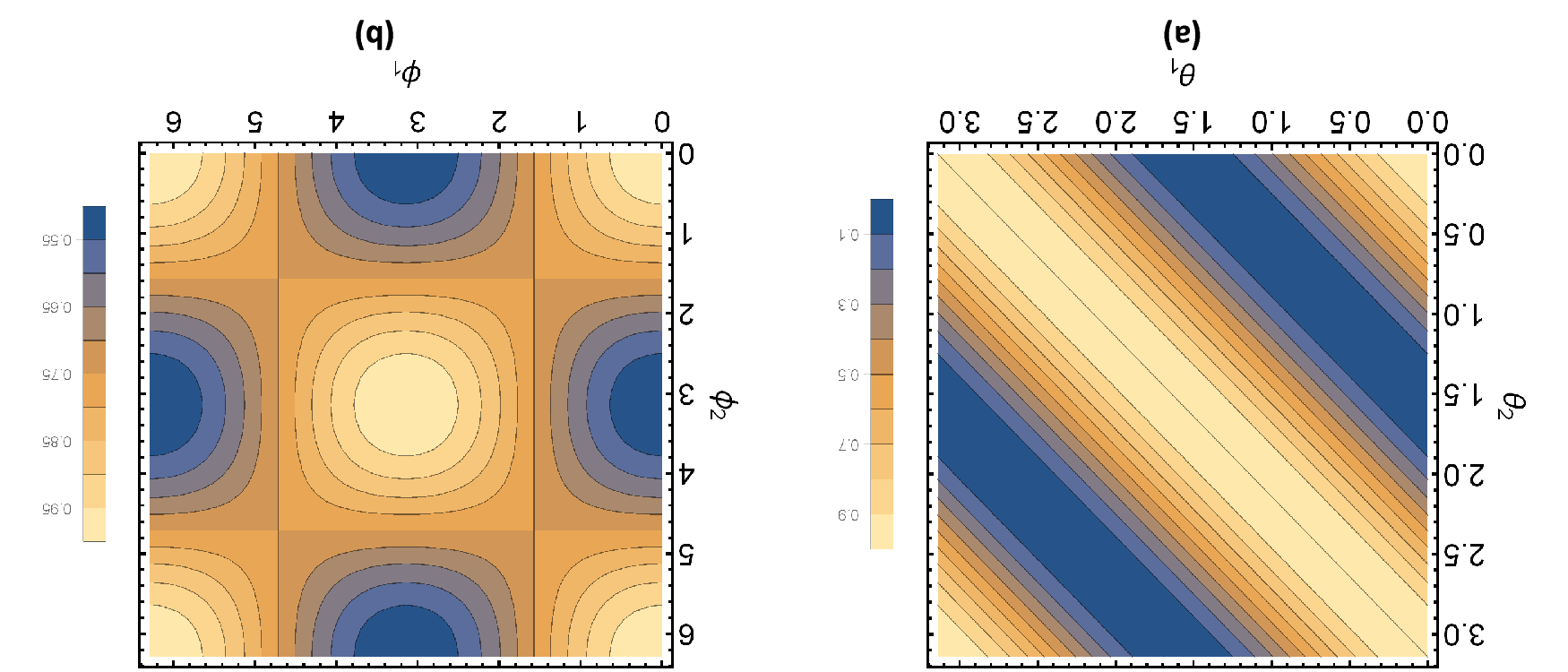}
\par\end{centering}

\protect\caption{\label{fig:QSDC-3D}Contour plots of fidelity of LM05 protocol when subjected to
CR and CD noises in (a) and (b), respectively.}
\end{figure}

Both the QSDC protocols when subjected to noise are affected to different
extent. Precisely, as observed in the protocols discussed so far,
the single-qubit-based schemes have been found to be more efficient
as compared to entangled-state-based schemes in AD and PD noisy channels. This is also observed here
in Figs. \ref{fig:QSDC-2D} a and b. Under the assumption of the same noise
parameter for CD and CR noise for Alice to Bob and Bob to Alice
travel of the qubits, PP protocol is affected by the CR and
CD noise in a manner similar to the entangled-state-based QKA protocol.
In fact, in the entangled-state-based QKA
scheme one of the parties sends the raw key by PP type QSDC while
the other party announces it. Therefore, the effect of noise is the
same as in PP protocol. The single-qubit-based scheme has different
nature in Fig. \ref{fig:QSDC-2D} c and d as compared to the corresponding
QKD and QKA protocols. This can be attributed to the two way quantum
communication associated in this scheme, unlike the last two cases
where it was unidirectional. Further, in the presence
of CD noise, the benefit of bidirectional communication can be easily
observed as the observed fidelity is more than the previous cases.
The single qubits (in LM05) perform better when subjected
to CD noise, but suffer more under the influence of CR noise. In Fig. \ref{fig:QSDC-3D},
we have not shown the contour plots for the fidelity for PP protocol under collective
noise as the expressions are exactly the same as that illustrated through Fig. \ref{fig:QKA-3D} for Shukla et al's QKA scheme. The contour plots also show that very low
fidelity is also possible for some particular values of noise parameters
during the two directions of transmission. Further, under the effect
of CD noise, a similar nature of the fidelity variation under LM05
and PP protocols can be observed in Fig. \ref{fig:QSDC-3D}. However,
a closer look reveals that under CD noise fidelity obtained for LM05 protocol is
more than that obtained for PP protocol, indicating that for CD noise, single-qubit-based LM05 protocol performs better than the corresponding entangled-state-based PP protocol. 

The expressions of fidelity under Pauli noise reveal that the
fidelity for PP protocol in bit, phase and bit-phase flip is the same
as the fidelity for bit-phase flip errors for equal probability of
error in LM05 protocol. In this case the fidelity resurrects to 1 for
maximum probability of error. Quite a similar nature is observed for
fidelity under bit flip and phase flip errors in LM05 scheme though
it remains less than that of the corresponding values in the PP protocol.
Similarly, under the influence of the depolarizing channel the fidelity fails to revive but
remains always more for LM05 protocol. 

The advantage of squeezing, a purely quantum resource, can be observed
in Fig. \ref{fig:QSDC-2D} f, where in the presence of squeezing after
an appreciable amount of time, fidelity higher than the corresponding case of zero squeezing
can be observed. Specifically, higher fidelity under SGAD channel
relative to AD channel shows that coherence can be sustained using
squeezing that would have been lost due to the presence of non-zero
temperature.

\subsection{Quantum dialogue protocols and effect of noise on them}

One of the most efficient secure quantum communication schemes is the quantum dialogue (QD). In this scheme both the legitimate
parties encode their information on the same qubits and at the end
of the protocol each party can deduce the others message. The first
QD scheme was proposed by Ba An using Bell states in 2004 \cite{ba-an}.
Recently, Yang and Hwang proposed a QD scheme immune to the collective
noise using logical qubits \cite{QD-hwang}. Here, we consider two QD protocols
for analyzing their performance when subjected to noisy environments.

\subsubsection{Single-qubit-based QD protocol}

A modified QD protocol using only single qubit states and MUBs was
proposed by Shi et al. in 2010 \cite{QD-singlepar}. Shi et al.
protocol can be described in the following steps:
\begin{description}
\item [{QD~1}] Bob prepares a sequence of $2n$ single qubits randomly in
$\left\{|0\rangle,|1\rangle\right\} $ or $\left\{|+\rangle,|-\rangle\right\} $ basis. He intentionally
prepares two copies of $n$ qubits, i.e., two adjacent photons are in
the same quantum state. For example, Bob prepares a string of single qubits as $\left\{\left(|0\rangle,|0\rangle\right),\left(|+\rangle,|+\rangle\right),\left(|-\rangle,|-\rangle\right),\left(|1\rangle,|1\rangle\right)\right\}$, and out of each pair, one qubit will be used for encoding while the other one will be used to send the initial state information. He also prepares some additional decoy qubits
to be used for eavesdropping check in each round of communication.
Finally, he sends all the $2n$ qubits after inserting decoy qubits
randomly in them to Alice.
\item [{QD~2}] Bob and Alice perform security checking for the received
qubits. Specifically, Bob will announce positions of the decoy qubits
he chooses for this round of communication and Alice announces her
choice of measuring basis and corresponding outcome. Using that Bob
determines the error rate and decides whether to proceed or call off
the protocol.
\item [{QD~3}] With the help of Bob, Alice can separate three sequences:
the first one of decoy qubits, and two sequences of one copy of initial
states each. Out of these three sequences she encodes her message
on the second sequence using $I$ ($iY$) operation for sending bit
value 0 (1). Subsequently, she encodes a checking message on the decoy
qubits using the same scheme and concatenates these two encoded sequences.
Finally, she sends this concatenated sequence after randomizing to
Bob while keeping the last sequence with herself.
\item [{QD~4}] After receiving authenticated receipt of all the qubits from Bob,
Alice will announce the positions of the decoy qubits and being aware
of the preparation basis he decodes the message and announces it publicly.
With this checking message Alice decides whether to go to the next step
or start afresh.
\item [{QD~5}] If they decide to proceed, Bob also encodes on the received
qubits after rearranging them using the same encoding scheme as Alice.
Subsequently, he performs measurement on all these qubits in the
basis in which they were prepared and announces the measurement outcomes. From
the measurement outcomes, Bob gains the knowledge of Alice's encoding
as he knows his encoding apart from the initial and final state. Further,
the measurement outcomes also reveal the choice of the basis used for
preparation of the states. Using this information Alice measures the
third sequence, she had kept with herself in QD 3, in a suitable basis
and learns the initial state of Bob. From the information of the initial
and final states, Alice can extract the message of Bob by using her
knowledge of her encoding.
\end{description}
The fidelity expression of single-qubit-based QD scheme under AD
channel contains cubic terms
\begin{equation}
F_{AD1}^{QD}=\frac{1}{8}\left(-2\eta^{3}+5\eta^{2}-\left(\sqrt{1-\eta}+7\right)\eta+2\left(\sqrt{1-\eta}+3\right)\right),
\end{equation}
which signify bidirectional quantum communication apart from a QSDC
to inform Bob about the initial state. This fact can also be observed
in PD noise

\begin{equation}
F_{PD1}^{QD}=\frac{1}{8}\left(-\eta^{3}+4\eta^{2}-6\eta+8\right).
\end{equation}
For the two rounds of communication under collective noisy channels, characterized
by two noise parameters, the fidelity for CD is

\begin{equation}
F_{CD1}^{QD}=\frac{1}{8}\left(\cos^{2}(\text{\ensuremath{\phi}}_{1})\cos(\text{\ensuremath{\phi}}_{2})+\cos(\text{\ensuremath{\phi}}_{1})
(\cos(\text{\ensuremath{\phi}}_{2})+1)+5\right),
\end{equation}
and for CR noise is

\begin{equation}
F_{CR1}^{QD}=\cos^{2}(\text{\ensuremath{\theta}}_{1})\cos^{2}(\text{\ensuremath{\theta}}_{1}+\text{\ensuremath{\theta}}_{2}).
\end{equation}
Similarly, all the qubits traveling through a Pauli
channel give rise to the fidelity as a function of various parameters, as

\begin{equation}
\begin{array}{lcl}
F_{P1}^{QD} & = & \frac{1}{2}\left\{ 2p_{1}^{3}+3p_{1}^{2}(p_{2}+p_{4})+2p_{1}\left(2p_{2}^{2}+p_{2}p_{3}+p_{3}^{2}+p_{3}p_{4}+2p_{4}^{2}\right)\right.\\
 &  & \left.+p_{2}^{3}+p_{2}^{2}p_{4}+p_{2}\left(p_{3}^{2}+4p_{3}p_{4}+p_{4}^{2}\right)+p_{4}\left(p_{3}^{2}+p_{4}^{2}\right)\right\}.
\end{array}
\end{equation}
Finally, the obtained fidelity for this QD scheme under the action of the SGAD channel is

\begin{equation}
\begin{array}{lcl}
F_{SGAD1}^{QD} & = & \frac{1}{16}\left\{ Q^{3}\left(-4\lambda^{3}+4(3\mu+\nu)\lambda^{2}-2\left(6\mu^{2}+4\nu\mu+2\nu^{2}+\sqrt{1-\lambda}-3\sqrt{1-\mu}\sqrt{1-\nu}\right)\lambda\right.\right.\\
 & - & 6\sqrt{1-\lambda}\mu+9\sqrt{1-\lambda}\mu\nu-6\sqrt{1-\lambda}\nu-5\sqrt{1-\mu}\mu\sqrt{1-\nu}\nu+2\sqrt{1-\mu}\sqrt{1-\nu}\nu\\
 & + & \left.4(\mu+\nu)\left(\mu^{2}+\nu^{2}\right)+8\sqrt{1-\lambda}+2\sqrt{1-\mu}\mu\sqrt{1-\nu}-8\sqrt{1-\mu}\sqrt{1-\nu}\right)\\
 & - & Q^{2}\left(12\mu^{3}+2(6\nu-5)\mu^{2}+3\left(\nu\left(4\nu+6\sqrt{1-\lambda}-5\sqrt{1-\mu}\sqrt{1-\nu}-5\right)+2\sqrt{1-\mu}\sqrt{1-\nu}\right)\mu\right.\\
 & + & 2\left(-6\sqrt{1-\lambda}\mu+\mu+\nu+2\sqrt{1-\lambda}\left(\sqrt{1-\mu}\sqrt{1-\nu}-3\nu\right)+6\sqrt{1-\lambda}-6\sqrt{1-\mu}\sqrt{1-\nu}-2\right)\\
 & + & 2\nu\left(\nu(6\nu-5)+3\sqrt{1-\mu}\sqrt{1-\nu}\right)+2\lambda\left(6\nu-2\left(2\nu^{2}+4\mu\nu+\mu(6\mu-5)\right)+3\sqrt{1-\mu}\sqrt{1-\nu}+1\right)\\
 & + & \left.2\lambda^{2}(6\mu+2\nu-5)\right)+\left(3\left(\nu\left(4\nu+3\sqrt{1-\lambda}-5\sqrt{1-\mu}\sqrt{1-\nu}-10\right)+2\sqrt{1-\mu}\sqrt{1-\nu}\right)\mu\right.\\
 & + & 12\mu^{3}+2\left(\left(8-3\sqrt{1-\lambda}\right)\mu+8\nu+\sqrt{1-\lambda}\left(2\sqrt{1-\mu}\sqrt{1-\nu}-3\nu\right)+4\sqrt{1-\lambda}-4\sqrt{1-\mu}\sqrt{1-\nu}-2\right)\\
 & + & \left.\left.4(3\nu-5)\mu^{2}-4\lambda\left(3\mu^{2}+(2\nu-5)\mu+(\nu-3)\nu+3\right)+2\nu\left(2\nu(3\nu-5)+3\sqrt{1-\mu}\sqrt{1-\nu}\right)\right)\right\}.
\end{array}
\end{equation}

\subsubsection{Ba An protocol of QD}

In the originally proposed Ba An's QD scheme, both parties can communicate
simultaneously using Bell states \cite{ba-an,baan_new,qd}. The protocol
can be summarized in the following steps:
\begin{description}
\item [{QD:BA~1}] Bob prepares $|\psi^{+}\rangle^{\otimes n}:\,|\psi^{+}\rangle=\frac{|00\rangle+|11\rangle}{\sqrt{2}}$.
He encodes his message on the first qubit (travel qubit) and keeps
the second qubit with himself as home qubit. To encode his message
he uses dense coding, i.e., he applies unitary operations $I,\,X,\,iY$
and $Z$ to encode $00,\,01,\,10$ and $11$, respectively. 
\item [{QD:BA~2}] Bob sends all the first qubits to Alice and confirms
their receipt. 
\item [{QD:BA~3}] Alice also encodes on the travel qubit using the same rule
as was used by Bob and sends them back to Bob. Bob performs a Bell measurement
on the partner particles (Bell measurement is done on a qubit from the sequence of home  qubits and another qubit from the sequence of travel qubits, which was initially entangled with the chosen home qubit). 
\item [{QD:BA~4}]  After Alice's disclosure Bob comes to know whether it
was message mode (MM) or control mode (CM)\footnote{In fact, a random choice of MM or CM mode by Alice provides security in the protocol. In CM mode, both the legitimate parties opt to check eavesdropping while in MM mode they proceed with the communication.}. Bob announces his measurement
outcome in the MM using which both Alice and Bob can learn each others
message. While in CM Alice announces her encoding which Bob uses for
eavesdropping checking. 
\end{description}
In the original Ba An's QD scheme, when subjected to AD and PD noise,
the fidelity can be seen to be
\begin{equation}
F_{AD2}^{QD}=\frac{1}{4}(\eta-2)^{2},
\end{equation}
and

\begin{equation}
F_{PD2}^{QD}=\frac{1}{2}\left(\eta^{2}-2\eta+2\right),
\end{equation}
respectively. The presence of quadratic terms is a signature of
bidirectional quantum communication involved. Under the coherent effect
of CD noise on the travel qubits, we obtain

\begin{equation}
F_{CD2}^{QD}=\frac{1}{2}\left\{ \cos(\text{\ensuremath{\phi}}_{1})\cos(\text{\ensuremath{\phi}}_{2})+1\right\}, 
\end{equation}
and for CR noise the fidelity is found to be

\begin{equation}
F_{CR2}^{QD}=\frac{1}{2}\left\{ \cos^{2}(\text{\ensuremath{\theta}}_{1}-\text{\ensuremath{\theta}}_{2})+\cos^{2}(\text{\ensuremath{\theta}}_{1}+
\text{\ensuremath{\theta}}_{2})\right\}.
\end{equation}
When the travel qubit is transmitted through
a Pauli channel the fidelity is the same as that obtained in case of
PP protocol. The analytic expression of fidelity for Ba An
protocol of QD, when subjected to SGAD noise is

\begin{equation}
\begin{array}{lcl}
F_{SGAD2}^{QD} & = & \frac{1}{4}\left\{ Q^{2}\left(\lambda^{2}-2\lambda(\mu+\nu+1)-2\left(2\sqrt{1-\lambda}\sqrt{1-\mu}\sqrt{1-\nu}+\mu+\nu-2\right)+\mu^{2}+5\mu\nu+\nu^{2}\right)\right.\\
 & + & \mu^{2}+\mu(5\nu-4)+(\nu-2)^{2}+\mu\nu(Q-1)^{2}\cos(2\Phi)\\
 & + & \left.2Q\left(\lambda(\mu+\nu-1)+2\sqrt{1-\lambda}\sqrt{1-\mu}\sqrt{1-\nu}-\mu^{2}+\mu(3-5\nu)-(\nu-3)\nu-2\right)\right\} .
\end{array}
\end{equation}

\begin{figure}
\begin{centering}
\includegraphics[scale=0.70,angle=0]{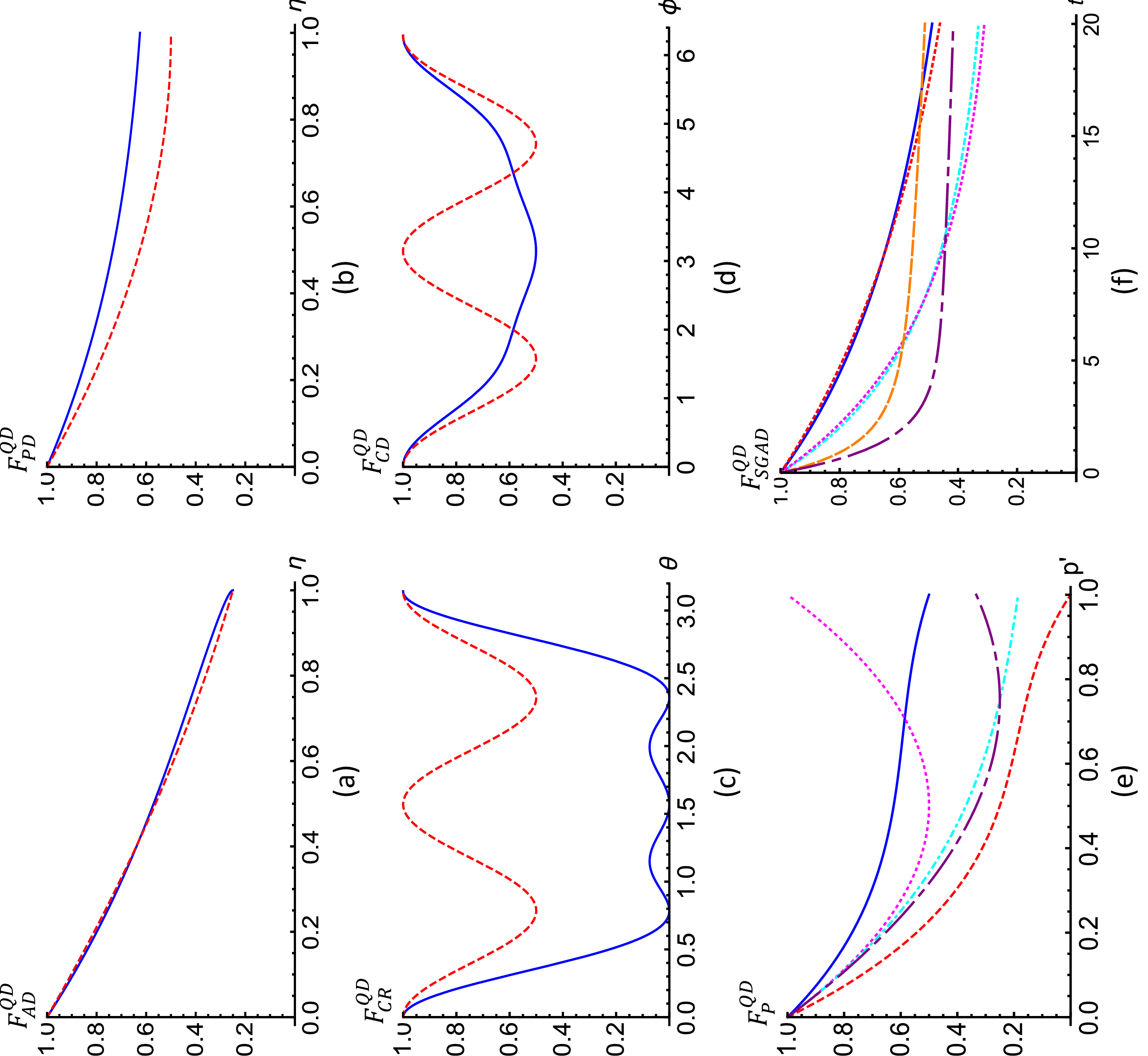}

\protect\caption{\label{fig:QD-2D}QD protocols are analyzed under AD, PD, CR and CD noises in (a)-(d), respectively. The smooth (blue) and dashed (red) lines correspond
to the single-qubit-based and Ba An's QD protocols, respectively. For
CR and CD noises it is assumed that the noise parameter is same for
both the directions of travel of the qubit (i.e., Alice to Bob and
Bob to Alice). In (e), bit/phase flip is shown together
for the single-qubit-based QD and Ba An protocol of QD in smooth (blue) and
dotted (magenta) lines, respectively. For Ba An protocol of QD bit phase
flip errors matches exactly with the previous case. However, for single-qubit-based scheme, it is shown in dashed (red) line. Under the
depolarizing channel the fidelity variation for the single-qubit-based scheme and Ba An protocol of QD is demonstrated by dotted dashed (cyan)
and large dotted dashed (purple) lines, respectively. (f)
corresponds to the effect of AD (i.e., an interaction with a zero
temperature and squeezing bath) in smooth (blue) and dashed (red)
lines; GAD (i.e., an interaction with a non-zero temperature and zero
squeezing bath) in dotted dashed (cyan) and dotted (magenta) lines
with $T=1$; and SGAD (finite temperature and squeezing bath) in large
dashed (orange) and large dotted dashed (purple) lines with $T=1$
and squeezing parameters $r=1$ and $\Phi=\frac{\pi}{8}$ for the
single-qubit-based and Ba An's QD protocols, respectively.}
\par \end{centering}
\end{figure}

\begin{figure}
\begin{centering}
\includegraphics[angle=270]{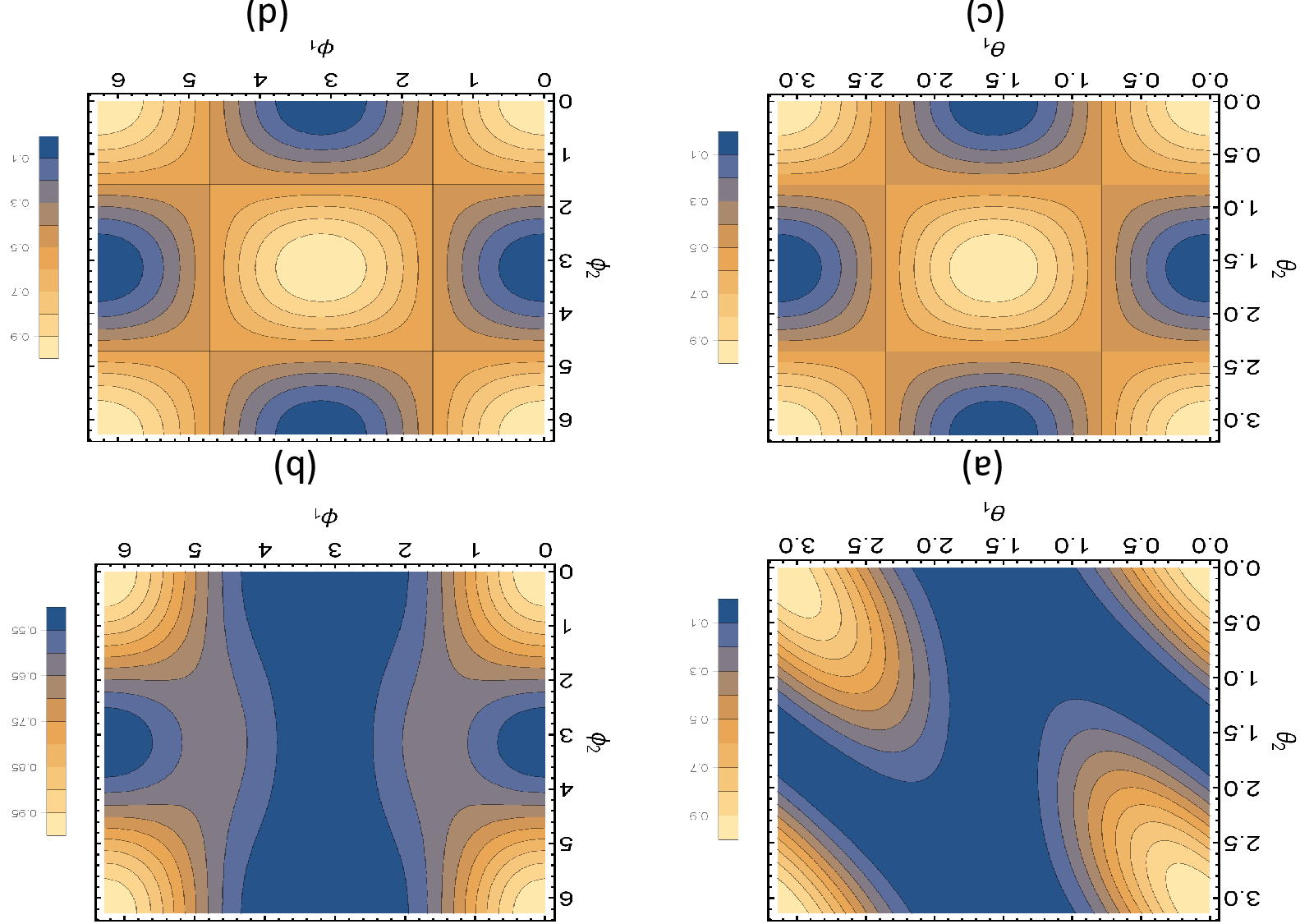} 
\par\end{centering}

\protect\caption{\label{fig:QD-3D}Contour plots illustrating the variation of fidelity of the single-qubit-based Shi et al. protocol and entangled-state-based Ba An protocol of QD in CR and CD noises, respectively.}
\end{figure}

When
both the protocols of QD are subjected to AD noise the fidelity obtained for the entangled-state-based protocol is comparable of that of the single-qubit-based one. This is in contrast with the earlier observations reported in the present work, where single-qubit-based schemes were found to be preferable in cases of AD and PD noise models.
Though, in the large decoherence limits the single-qubit-based
QD turns out to be a suitable candidate (cf. Fig. \ref{fig:QD-2D} a). It is worth
commenting here that the decay in the fidelity of single-qubit-based
QD scheme when compared with the corresponding QSDC protocol (as they
are of the same order in entanglement-based schemes) can be attributed
to an extra single qubit traveling through the noisy channel in step
QD 1. However, under the effect of PD channels, the observation established from the previous three
secure quantum communication schemes (namely, QKD, QKA and QSDC protocols) remains valid (cf. Fig. \ref{fig:QD-2D} b), in other words, it is observed that the single-qubit-based
schemes perform  better in PD channels. When Ba An protocol of QD is subjected to collective noise the same nature
of fidelity variation as was observed in PP protocol is observed if the same noise parameters
are used in to and fro travel of the qubits. However, in the single-qubit-based QD scheme a different nature from LM05 protocol is observed.
Interestingly, a close look at Figs. 
\ref{fig:QSDC-2D} c (d), and Fig. \ref{fig:QD-2D} c (d) reveals
that compared to LM05 protocol, an extra dip is observed at $\theta=\frac{\pi}{2}$
(where for LM05 fidelity was obtained to be unity). This dip was observed
in single-qubit-based QKA scheme, too (cf. Fig. \ref{fig:QKA-2D} c). This point further establishes
the fact that the fidelity of single-qubit-based
QD schemes decays, when subjected to AD noise. The contour plots shown in
Fig. \ref{fig:QD-3D} also demonstrate that the single-qubit-based
QD scheme has a different nature of fidelity variation compared to that in LM05 protcol, while in entangled-state-based protocol it remains similar to that observed in PP.    

The expression of fidelity under Pauli noise shows that for Ba An
protocol of QD it is the same as in PP protocol. However, the presence of
cubic terms in the expressions of fidelity for single-qubit-based
QD is a signature of the nature observed in Fig. \ref{fig:QD-2D} e,
i.e., the descent for very low and high error probabilities. For
the bit/phase flip error the single-qubit-based scheme remains the
preferred choice, but for very high probability of errors it should
be avoided. A similar nature is also observed under the influence of a depolarizing
channel. For bit-phase flip error, fidelity indicates better performance of
entanglement-based scheme as compared to their single-particle counterparts.

Under a dissipative interaction with a non-zero temperature bath, a behavior similar to that
observed under an AD channel is seen. However, due
to the non-zero squeezing, single-qubit-based QD scheme turns out
to be a better candidate. Fig. \ref{fig:QD-2D} f  further reiterates
the facts  observed in Fig. \ref{fig:QD-2D} a-d, i.e., the obtained
nature in the case of Ba An protocol of QD is the same as in PP protocol; and
in single-qubit-based scheme, it can be explained as a compound
effect of single-qubit-based QKA and QSDC protocols.

\section{Conclusion \label{sec:Conclusion}}

The comparative study of single-qubit-based and entangled-state-based
schemes of secure quantum communication performed in the present work
has lead to a number of interesting conclusions. Firstly, the equivalence
observed in the ideal noiseless scenario is lost in  more practical
scenarios where noise is present. Next, it is observed that it
is not possible to say unambiguously that in a noisy environment entangled-state-based
protocols perform better than the single-qubit-based protocols or vice versa. In fact, it depends on the nature of the noise present
in the channel. Specifically, single-qubit-based schemes are generally
found to be the suitable choice in the presence of AD and/or PD noises, while
entanglement-based protocols turns out to be preferable in the presence
of collective noises. As SGAD and GAD channels are generalizations
of the AD channel, conclusions similar to that for the AD channel are also applicable to them. 
However, with an increase in temperature, fidelity is seen to decay. Squeezing is seen to play the role of a beneficial quantum resource, 
in a host of scenarios, in practical quantum communication. Also, it is observed that
the effect of noise depends upon the number of rounds (how many times a travel qubit travels through the noisy channel) of quantum communication
involved. For instance, QKD protocols are least affected by noise,
while QD protocols are most affected as in QKD protocols a travel qubit travels only once through the noisy channel, whereas in Ba An protocol of QD, it travels twice through the noisy channel. Further, the single-qubit-based
QD scheme involves three rounds of communication as it requires Alice
to Bob and Bob to Alice transmission of qubits and an additional Alice
to Bob travel of  equal number of qubits. As a consequence, single-qubit-based
QD scheme is found to be the most affected among
the four different single-qubit-based schemes for secure quantum communication discussed in this paper. 

Acknowledgement: AP acknowledges the support provided by DST, India through the project number EMR/2015/000393.

\end{document}